\documentclass{article}
\usepackage{osa2}
\usepackage{epsfig}
\def\prn#1{{\left(#1\right)}}

\def\sbrk#1{{\left[#1\right]}}
\begin{document}

\vspace{1.cm}

\title{Variable Free Spectral Range Spherical Mirror Fabry-Perot Interferometer}
\author{Katherine Kerner$^{a}$, Simon M. Rochester$^{a}$,  Valeriy V. Yashchuk$^{a}$, and D. Budker$^{a,b,*}$}
\address{$^a$Department of Physics, University of California at Berkeley, Berkeley, CA 94720-7300
\\$^b$Nuclear Science Division, Lawrence Berkeley National Laboratory, Berkeley CA 94720}
\email{$^{*}$budker@socrates.berkeley.edu}

\centerline{ABSTRACT}

\begin{quotation}

A spherical Fabry-Perot interferometer with adjustable mirror
spacing is used to produce interference fringes with frequency
separation $(c/2L)/N$, $N=2-15$. The conditions for observation of
these fringes are derived from the consideration of the eigenmodes
of the cavity with high transverse indices.

\end{quotation}


\newpage
\vfill\eject

\vspace{1.cm}

\section*{Introduction}

A spherical mirror Fabry-Perot interferometer in the confocal
configuration has many advantages over plano-plano
interferometers, such as easier construction and alignment. These
interferometers are used in a variety of spectroscopic
applications, including laser spectrum analysis and generating
frequency markers for laser frequency scans. Here we describe how
one can employ a spherical mirror interferometer to produce
fringes with significantly smaller free spectral range
($FSR$=frequency interval between adjacent transmission peaks) by
adjusting the mirror separation to specific values different from
the confocal condition$^{1}$. This is useful when closely- or
variably-spaced frequency markers are required.

\section*{Theory}

\begin{figure}\begin{center}\epsfig{file=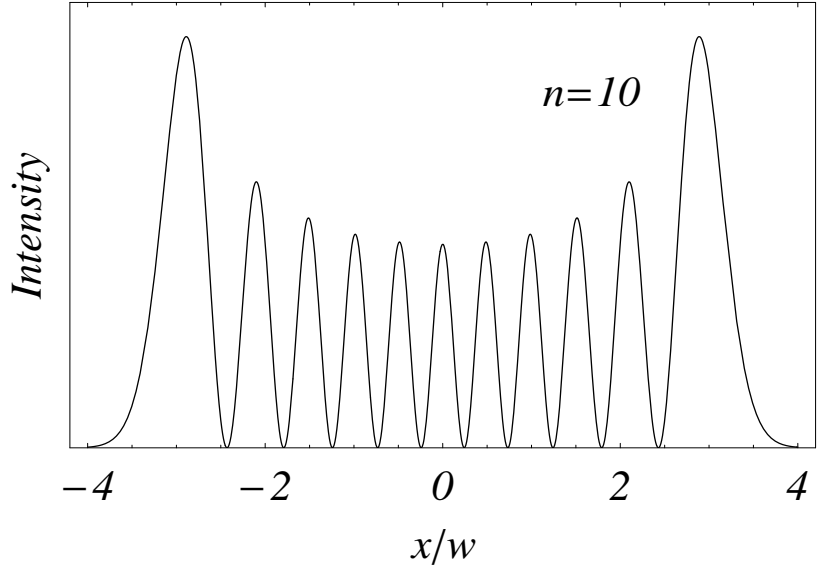,width=5 in,height=3 in}
\caption{Intensity profile for the Hermite-Gaussian mode pattern
with $n=10$, where $x$ is the transverse distance from the beam
center and $w$ is the $1/e$ spot size for the lowest order ($n=0$)
transverse mode.}\label{Intensity}\end{center}\end{figure}

Consider a symmetric optical resonator consisting of two spherical
mirrors with radius of curvature $R$, separated by a distance $L$
in the $\hat{z}$ direction$^{2}$. The optical modes in this
resonator are well approximated by the Hermite-Gaussian modes (in
the paraxial approximation), with $\hat{x}$ and $\hat{y}$
direction transverse mode numbers $n,m$ corresponding to the
number of null points in the transverse intensity profile (Fig.
\ref{Intensity}). In free space, a Gaussian beam with transverse
mode numbers $n,m$ experiences an additional phase shift in
passing through a focal region of $(m+n+1)\pi$ (the Gouy phase
shift) relative to a plane wave. In a resonator, the beam is
confined to a finite region around the focal point, so that the
total Gouy phase shift is reduced. In this case, the phase shift
experienced by the beam with wavelength $\lambda$ in a double pass
(of distance $2L$) of the resonator is
\begin{equation}
\varphi(k=m+n)=2\pi\frac{2L}{\lambda}-2(k+1)\arccos
\sbrk{1-\frac{L}{R}} , \label{phase}
\end{equation}
where the first term represents the phase shift that would be
experienced by a plane wave, and the second term is the total Gouy
phase shift inside the resonator. The resonance condition is that
the round trip phase shift $\varphi(k)=2\pi q$, where $q$ is an
integer (the axial mode number). Substituting this condition into
Eq. \ref{phase}, gives the resonance frequencies:
\begin{equation}
\nu(q,k)=\frac{c}{2L}\prn{q+
\frac{k+1}{\pi}\arccos\sbrk{1-\frac{L}{R}}}. \label{Frequency}
\end{equation}

\begin{figure}\begin{center}\epsfig{file=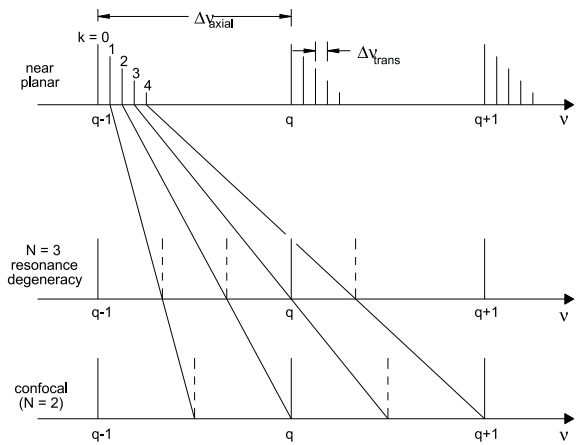,scale=0.6}
\caption{Relative positions of the lowest order transverse modes
shown for the cases of a near-planar ($L\ll R$), and confocal
($L=R$) interferometers, as well as for the degeneracy condition
where $N=3$. }\label{TransverseModes}\end{center}\end{figure}

Without mode-matching (i.e. preferential coupling into a single
mode), one typically excites many transverse modes of the
interferometer. For example, for the interferometer used in the
present work ($R=2.5$ cm, $L\approx2.5$ cm, $\lambda=780$ nm), the
size of the fundamental transverse mode on a mirror is
$w\approx\prn{L\lambda/\pi}^{1/2}\approx 0.1$ mm. Since the width
of higher transverse modes with index $m$ is roughly $w\sqrt{m}$,
if we illuminate the input mirror with a laser beam, for example,
of a width $\sim1$ mm, we expect that transverse modes with
$m<m_{max}\sim100$ will be excited (and even higher modes if the
beam is offset from the axis). Similarly, we have $n<m_{max}$ for
the other transverse direction. As Eq. \ref{Frequency} indicates,
the frequencies of the transverse modes generally do not coincide
with those of the axial modes, producing a complex and irregular
pattern of fringes at the output as a function of the input laser
frequency. As the mirror separation is changed, the frequencies of
the transverse modes move with respect to those of the axial modes
(Fig. \ref{TransverseModes}). In the confocal configuration (where
$L=R$), every other transverse mode becomes degenerate with an
axial mode, producing a pattern of fringes with half the axial
spacing ($FSR=c/(4L)$). This removes the need for mode matching
and is one of the reasons a confocal interferometer is
particularly useful$^{3}$.

If the mirror separation is adjusted away from the confocal
condition, we can find higher-order degeneracies where every
$N^{th}$ transverse mode is degenerate with an axial mode, i.e.
$\nu(q,k+N)=\nu(q+l,k)$, where $N,l$ are integers. The output
fringes in this case have smaller $FSR=c/(2LN)$. From Eq.
\ref{Frequency}, one obtains the resonance conditions
\begin{equation}
\frac{L}{R}= 1-
\cos\sbrk{\frac{l\pi}{N}};~~l,N\mbox{~mutually~prime,} ~l<N.
\label{Resonance}
\end{equation}
The additional conditions are added to avoid double counting
resonances with higher degeneracies.

\begin{figure}\begin{center}\epsfig{file=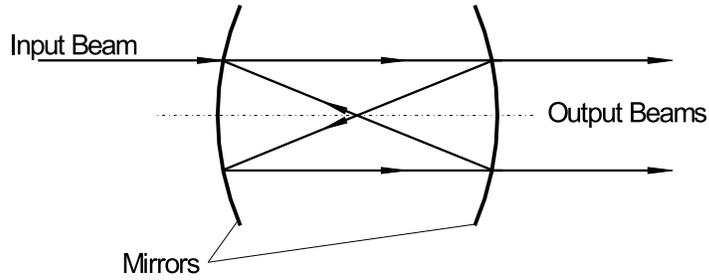,scale=0.6}
\caption{A schematic of the ray propagation in the confocal
configuration. Note that the ray traverses four times before
overlapping with itself and is transmitted in two places,
producing two spots on the output mirror ($N=2$)
.}\label{Confocal}\end{center}\end{figure}

The appearance of resonances corresponding to different values of
$N$ may also be understood in the ray-tracing approach applied in
Ref. 4. In that work, the relation of the extra cavity resonances
to the appearance of closed ray paths and applications to laser
resonators and absorption cells are discussed. For an arbitrary
mirror separation, a light ray (coming into the cavity off-axis)
never overlaps with its original location on the mirror. However,
in the confocal configuration, the beam returns to its original
position after traversing the cavity four times (Fig.
\ref{Confocal}). Two spots may usually be observed where the beam
hits the output mirror and is partially transmitted. At certain
mirror spacings (given by Eq. \ref{Resonance}), different from the
confocal separation, the beam returns to its original position
after making more than four traversals. In this situation, $N$
spots are observed, where $N$ corresponds to the resonance number
described above.

\section*{Apparatus}

\begin{figure}\begin{center}\epsfig{file=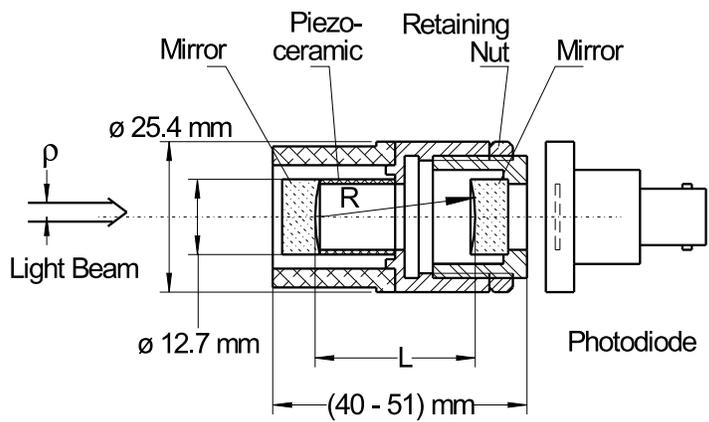,scale=0.6}
\caption{Cross-section of the confocal Fabry-Perot interferometer.
}\label{interferometer}\end{center}\end{figure}

In recent years, inexpensive home-made confocal devices have found
broad application both in research and instruction laboratories.
In our design$^{5}$ (Fig. \ref{interferometer}), the body of the
interferometer is constructed of two fine-threaded metal pipes.
One pipe threads into the other allowing accurate adjustment of
the mirror separation. A standard concave mirror (intensity
reflectivity $\mathcal{R}=95-98\%$, radius of curvature $R=2.5$
cm) is glued directly to one of the pipes. The second identical
mirror is glued to a piezo-ceramic hollow cylinder, which, in
turn, is glued to the second metal pipe. Application of voltage
between the walls of the piezo-ceramic tube displaces the mirror,
providing frequency tuning of the interferometer, typically, by
several free spectral ranges per 100 V. While scanning the
interferometer and observing the transmission fringes, one adjusts
the average mirror separation to achieve the confocal (or
higher-order degeneracy) condition (tolerance $\sim10^{-2}$ mm),
where the width of the observed transmission peaks is minimal and
their amplitude is at a maximum. Once the desired separation
between the mirrors is found, the spacing may be fixed by
tightening the retaining nut.

\section*{Experimental Results}

\begin{figure}\begin{center}\epsfig{file=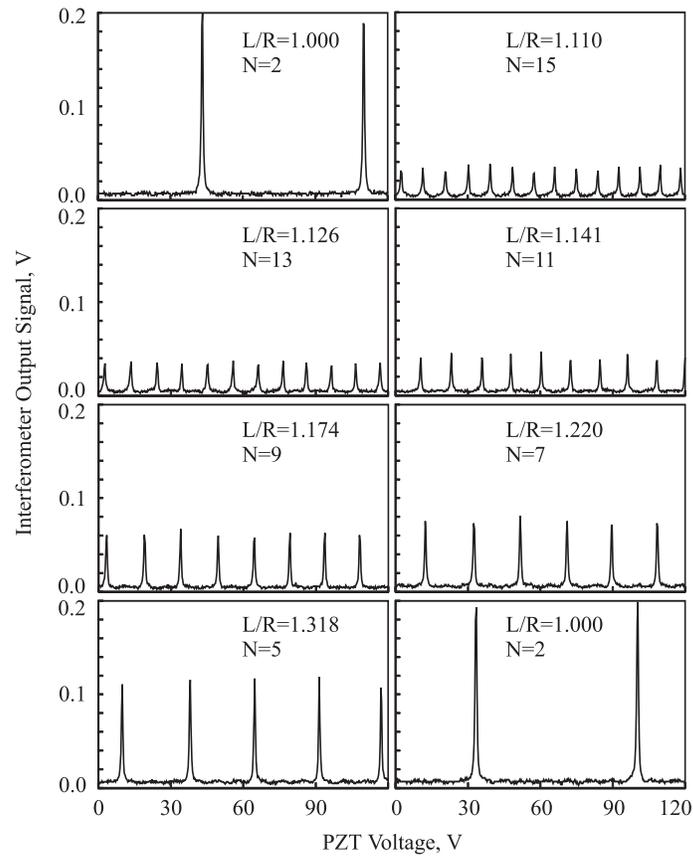,scale=0.6}
\caption{Transmission fringes recorded at various separations
corresponding to different $N$. The figure shows a single series
of measurements, and the device was returned to the confocal
configuration to check
reproducibility.}\label{experiment}\end{center}\end{figure}

\begin{figure}\begin{center}\epsfig{file=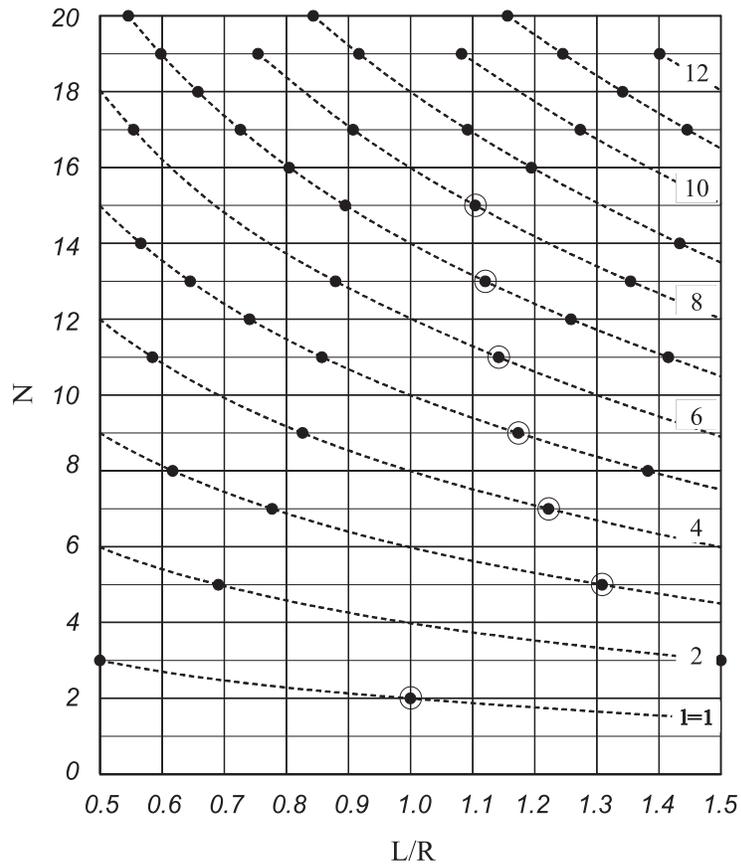,scale=0.6}
\caption{Comparison of the predicted (dots) and experimental
(circles) values of $L/R$ corresponding to degeneracies with
various values of $N$ and $l$ (see text).
}\label{Solutions}\end{center}\end{figure}

We have investigated the properties of the device described above
at mirror separations different from the confocal separation.
Narrow-band light from a commercial diode laser at $\lambda=780$
nm was directed into the interferometer. The mirror separation was
scanned by applying voltage to the piezo-ceramic cylinder, and the
output fringes were observed. The average mirror separation was
adjusted to values corresponding to resonances given by Eq.
\ref{Resonance}. We were able to produce well-resolved fringes
with $N$ up to 15 (Fig. \ref{experiment}), and measured the mirror
spacings corresponding to the resonances. Since a smaller number
of modes are degenerate between each other for $N>2$ compared to
the confocal case, the peak transmission of the interferometer is
reduced, roughly as $2/N$. The widths of the transmission peaks
corresponding to a given mode are mostly determined by the
mirrors' reflectivity and do not change with $N$. Thus, the
effective finesse (the ratio of $FSR$ to the transmission peak
width at half maximum) of the device also scales as $2/N$.
However, the decrease in $FSR$ as $N$ increases results in fringes
with small adjustable $FSR$, which allows the use of a single
compact device in place of multiple interferometers of much
greater length. The measured mirror separations corresponding to
the resonances coincided with the prediction of the theory of Eq.
\ref{Resonance} within experimental uncertainty of $\sim7$
microns. The predicted and measured values of $L/R$ are shown in
Fig. \ref{Solutions}.

\begin{figure}\begin{center}\epsfig{file=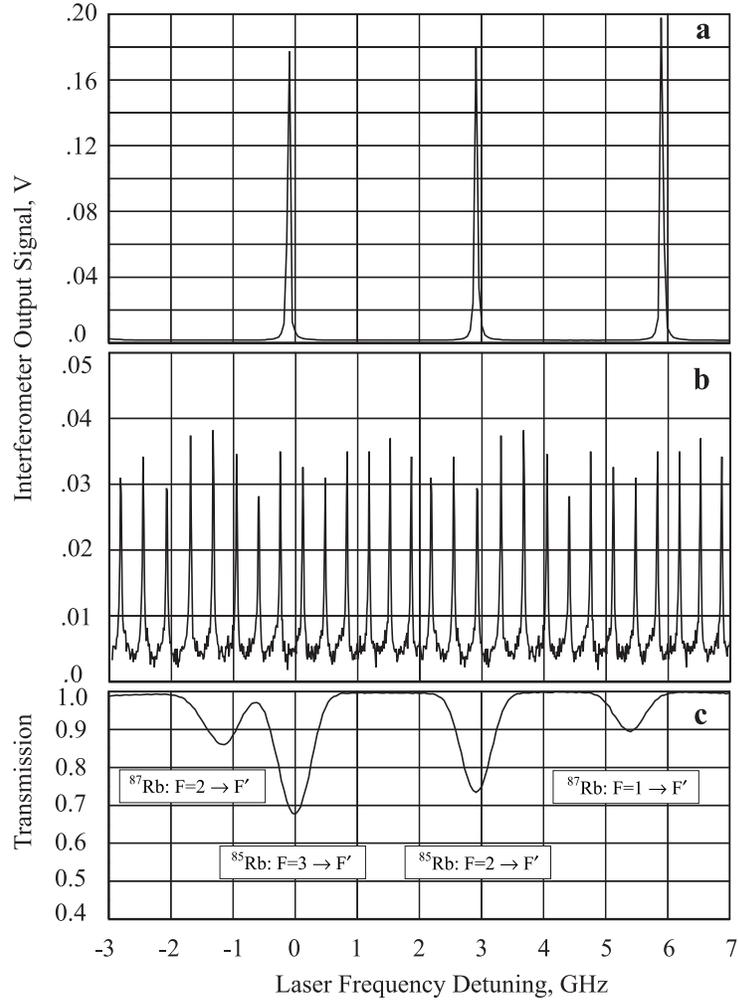,scale=0.6}
\caption{a.) Transmission fringes for interferometer in the
confocal regime ($L/R=1.00$) b.) Same for mirror separation
adjusted to $L/R=1.11$), corresponding to a resonance with $N=15$.
c.) Doppler-broadened transmission spectrum of the Rb D2-line
($\lambda=780$ nm) recorded with a low-power tunable diode laser
light passed through a room-temperature buffer-gas free vapor cell
($5$ cm long).}\label{example}\end{center}\end{figure}

Examples of experimental transmission patterns of an
interferometer with mirror separation in the vicinity of a
confocal resonance and a resonance with $N=15$ are given in Figure
\ref{example}. This Figure also shows the Doppler-limited
absorption spectrum of the Rb D2 line (Fig. \ref{example}c). While
the confocal fringes are adequate as frequency markers for the
Doppler-broadened scan, the higher order fringes are useful for
higher resolution scans, e.g. when a scan extends over just one of
the four peaks shown in Fig. \ref{example}c. which is often the
case in Doppler-free spectroscopy$^{6,5}$.

This work has been supported by NSF, grant PHY-9733479 and by ONR,
grant N00014-97-1-0214.


\newpage
\vfill\eject


\section*{REFERENCES}

\vspace{1.cm}

\begin{description}

\item[$^{\ 1}$]
    D.~Budker, S.~Rochester, and V.~V.~Yashchuk, Rev. Sci. Instr. {\bf 71}(8),
    (2000).
\item[$^{\ 2}$]
    In this discussion we follow the approach given by A.~E.~Siegman. {\it Lasers},
    University Science Books, Mill Valley, California, 1986.
\item[$^{\ 3}$]
    As the transverse mode number increases, the paraxial
   approximation breaks down. In order to limit the order of the
   transverse modes excited, we require that
   $\rho^{4}/R^{3}<\lambda/100 $, where $\rho$ is a typical size
   characterizing the width and offset from the axis of the input
   laser beam. This means that $\rho$ is limited to $\sim1~mm$ for
   the interferometers used in this work. Larger beam sizes and
   offsets lead to broader and asymmetric transmission peaks.
\item[$^{\ 4}$]
    D.~Herriott, H.~Kogelnik, and R.~Kompner, Appl. Opt. {\bf3}(4), 523(1964).
\item[$^{\ 5}$]
    D.~Budker, D.~J.~Orlando, and V.~Yashchuk, Am. J. Phys. {\bf
    67}, 584 (1999).
\item[$^{\ 6}$]
    W.~Demtr${\rm \ddot{o}}$der. {\it Laser specroscopy},
    Springer-Verlag, Berlin, Heidelberg, New-York, 1998.
\end{description}

\newpage
\vfill \eject

\newpage
\vfill \eject

\newpage
\vfill \eject

\newpage
\vfill \eject

\newpage
\vfill \eject

\newpage
\vfill \eject

\newpage
\vfill \eject

\end {document}